# Benford's Law, its applicability and breakdown in the IR Spectra of polymers

Ed. Bormashenko[*], E. Shulzinger, G. Whyman, Ye. Bormashenko

Ariel University, Faculty of Natural Sciences, Department of Physics, Ariel, P.O.B.3, 40700, Israel

ABSTRACT

Infrared spectra of various polymers were treated statistically. It was established that for the absorbance spectra the Benford distribution of leading digits takes place, whereas the distribution of leading digits for transmittance spectra is random. This observation may be explained by the fact that the value of transmittance $Tr$ is restricted $0 < Tr < 1$, due to the physical reasons, whereas the value of absorbance is not. Moreover, the transmittance and absorbance $A$ are interrelated by the logarithmic dependence $A = -\log Tr$. This observation supplies the idea that the Benford's law is valid in the situations, when logarithmic dependencies take place.

1. Introduction

A phenomenological, contra-intuitive law also called the first digit law, first digit phenomenon, or leading digit phenomenon, Benford's law states that in listings, tables of statistics, etc., the digit 1 tends to occur with probability ~30%, much greater than the expected 11.1% (i.e., one digit out of 9). The discovery of Benford's law goes back to 1881, when the American astronomer Simon Newcomb noticed that in logarithm tables (used at that time to perform calculations) the earlier pages (which contained numbers that started with 1) were much more worn and smudged than the later pages. S. Newcomb noted "that the ten digits do not occur with equal frequency must be evident to any making use of logarithmic tables, and noticing how much faster first pages wear out than the last ones" [1].

The phenomenon was re-discovered in 1938 by the physicist Frank Benford, who tested it on data extracted from 20 different domains, as different as the surface areas of rivers, the sizes of, physical constants, molecular weights, etc. [2]. Since that, the law is credited to Benford. The Benford law is expressed by the following statement: the occurrence of first significant digits in data sets follows a logarithmic distribution:

$$P(d) = \log_{10}(1+\frac{1}{d}), d = 1,2...9, \qquad (1)$$

where $P(d)$ is the probability of a number having the first non-zero digit $d$.

Since that, Benford's law was applied for the analysis of the statistical data related to a broad variety of statistical data, including atomic spectra [3], population dynamic [4], magnitude and depth of earthquakes [5], genomic data [6-7], mantissa distributions of pulsars [8] and economic data [9]. While Benford's law definitely applies to many situations in the real world, a satisfactory explanation has been given only recently through the works of Hill [10-12]. Important intuitive insights in the grounding of the Benford law, relating its origin to the scaling invariance of physical laws were reported recently by Pietronero [13]. Engel et al. demonstrated that the Benford law takes place approximatively for exponentially distributed numbers [14]. The breakdown of the Benford law was reported for certain sets of statistical data [15-16].

It should be mentioned that the grounding and applicability of the Benford law remain highly debatable [12]. Our paper focuses on the applicability of the Benford law for the analysis of infrared (IR) spectra of polymers. IR spectroscopy serves as one of the main instruments for the characterization and identification of polymers [17-18]. Every polymer has specific IR bands of transmission (absorbance). We applied the Benford law for the analysis of spectral data in two domains, namely transmission and absorbance spectra. The results were quite different. The spectral data in the absorbance domain follow the Benford law, whereas in the transmission domain they do not.

2. **Experimental**

IR transmission and absorbance spectra of following polymers: polyethylene, polypropylene, polystyrene, polycarbonate, polyethylene terephthalate, polyarylate

(Ardel), ethylene chlorotrifluoroethylene (Halar), non-polarized polyvinylidene fluoride (Kynar), polarized polyvinylidene fluoride, polyether ether ketone, polysulfone (Thermalux), polyethersulfone, polymethylpentene were taken. Consider that IR spectra of polarized and non-polarized spectra of polyvinylidene fluoride are quite different [18]. In addition, IR spectra of honeycomb polystyrene films, prepared as described in Ref. 19, were measured. In total, the 20 spectra of polymers were taken. The samples were extruded polymer films with the thickness of 25-50 μm.

Spectra were taken by the Bruker 22 FTIR spectrometer. A FTIR spectrometer Bruker 22 allowed spectral measurements in the fingerprint range of polymers, namely 400-2200 cm$^{-1}$ with the accuracy of 4 cm$^{-1}$. So every spectrum comprised a set of 934 points (approximately 2 points per every 4 cm$^{1}$). All spectral measurements were done at room temperature.

### 3. Results and discussion

IR spectra of very different polymers were studied including crystalline (polyethylene and polypropylene) and amorphous (polycarbonate, polysulfone, etc.), non-piezoelectric (non-polarized polyvinylidene fluoride) and polarized (polarized polyvinylidene fluoride) ones. The IR spectra of polymers were studied in two domains, namely the transmittance and absorbance ones. The transmittance at the certain wavelength $\lambda$ denoted as $Tr(\lambda)$ is defined as the ratio of the intensities of the transmitted light $I(\lambda)$ to the intensity of the incident light $I_0$ [20-21]:

$$Tr(\lambda) = \frac{I(\lambda)}{I_0} \qquad (2)$$

It is conventional in the spectroscopy to use wavenumbers $k = \frac{2\pi}{\lambda}$ instead of wavelengths, thus the transmittance spectrum of polymer is described by the function: $Tr(k) = \frac{I(k)}{I_0}$. Actually the spectroscopy measurement puts in the correspondence the set of discrete values of the transmittance to the set of wavenumbers, namely: $Tr_i(k_i) \to k_i$.

The absorbance *A,* characterizing the amount of light absorbed by the sample is defined, in turn, according to Eq. 3 [20-21]:

$$A(k) = -\log \frac{I(k)}{I_0} = \log \frac{1}{Tr(k)} \qquad (3)$$

Again, the absorbance spectrum puts in the correspondence the set of discrete values of the absorbance to the set of wavenumbers, i.e: $A_i(k_i) \to k_i$.

Now compare the spectral data obtained in the transmittance and absorbance domains from the point of view of the applicability of the Benford distribution, as shown in Figs. 1-3, depicting the data amalgamated from all spectra taken.

It is clearly seen from the data, supplied in Figs 1-3 that the distribution of the significant digits in the absorbance domain follows the Benford law (the first digit 1 appears about 35% times), whereas, in the transmittance domain the distribution of significant digits is random. Let us explain this observation. The researchers noted that the Benford distribution is self-consistent with the scaling invariance of studied physical systems [13]. If we analyze the stock data from the point of view of the Benford law it turns out that Benford's distribution (1) is independent of the currency adopted [13]. Pietronero et al. conjectured that the Benford law is applicable for a broad range of situations where the scaling invariance occurs [13].

It is noteworthy that the both of the absorbance and transmittance spectra demonstrate scaling invariance, this means that if we change the units of the wavenumbers (for example we may calibrate wavenumbers in ft$^{-1}$ instead of cm$^{-1}$), Eqs. 4, describing the spectra, take place:

$$Tr(k') = Tr(\alpha k) = \Psi(\alpha)Tr(k) \qquad (4a)$$

$$A(k') = A(\alpha k) = \Phi(\alpha)A(k) , \qquad (4b)$$

where $\alpha$ is the coefficient transferring $k$ to $k'$, $k' = \alpha k$ (for example cm$^{-1}$ to ft$^{-1}$) and $\Psi(\alpha), \Phi(\alpha)$ are functions depending on $\alpha$ but independent on $k$ [13]. It is noteworthy that in our case of the spectral data, experiments will supply the same sets of $A_i$ and $Tr_i$ irrespectively to the units of wavenumber $k_i$. This is the particular ("strong") case of scaling invariance discussed in Ref. 13.

Both of the absorbance and transmittance demonstrate the scaling invariance, however, Benford's distribution is valid for the absorbance domain only (see Figs. 2-3). The reasonable explanation of this observation is obtained by comparing formulae (2) and (3). The modulus of absorbance is the logarithm of transmittance, and it follows Benford's distribution, as it occurs for a broad range of logarithmic dependencies [1]. It also should be emphasized, that the value of transmittance is restricted due to obvious physical reasons, i.e. Eq. 5:

$$0 < Tr(k) = \frac{I(k)}{I_0} < 1 \qquad (5)$$

takes place. In these cases the Benford law is not applicable [16]. At the same time the value of absorbance is not restricted. It should be stressed that no Benford distribution is recognized for the spectra of specific polymers, as shown in Fig. 4, depicting the data extracted from the IR absorbance spectra of polycarbonate and polystyrene. This situation is typical, namely Benford's law works well for amalgamated data.

Now we conjecture, why the Benford distribution is so abundant in a broad variety of statistical data. The vast number of processes, occurring in the nature, are entropically driven. Entropy $S$ is defined, as $S = k \ln W$, where $k$ is the Boltzmann constant, and $W$ is the number of microstates corresponding to a given macrostate [22]. Thus, entropy always appears as a logarithm of a certain number. Perhaps, this is reason for the abundance of the Benford law in the statistical data.

4. **Conclusion**

We conclude that IR spectra of polymers supply the convenient object for studying the Benford law. The IR spectra may be treated as sets of values of transmittance and absorbance corresponding to the sets of wavenumbers. It was established that the Benford distribution works well in the domain of absorbencies, whereas it fails in the domain of transmittance, where the distribution of leading digits is random. We relate this observation to the fact that the value of transmittance is restricted, and $0 < Tr(k) < 1$ takes place. On the other hand, the absorbance $A$ is defined as: $A(k) = \log \frac{1}{Tr(k)}$, and this puts us back to the famous finding made by Newcomb with the use of logarithmic tables: the distribution of leading digits is not equal, when the logarithmic dependence is prescribed.

We conjecture that the abundance of the Benford law in physical data is related to the fact that entropy driven processes are widespread in nature, and the entropy, in turn, is the logarithm of the number of microstates corresponding to a given macrostate.

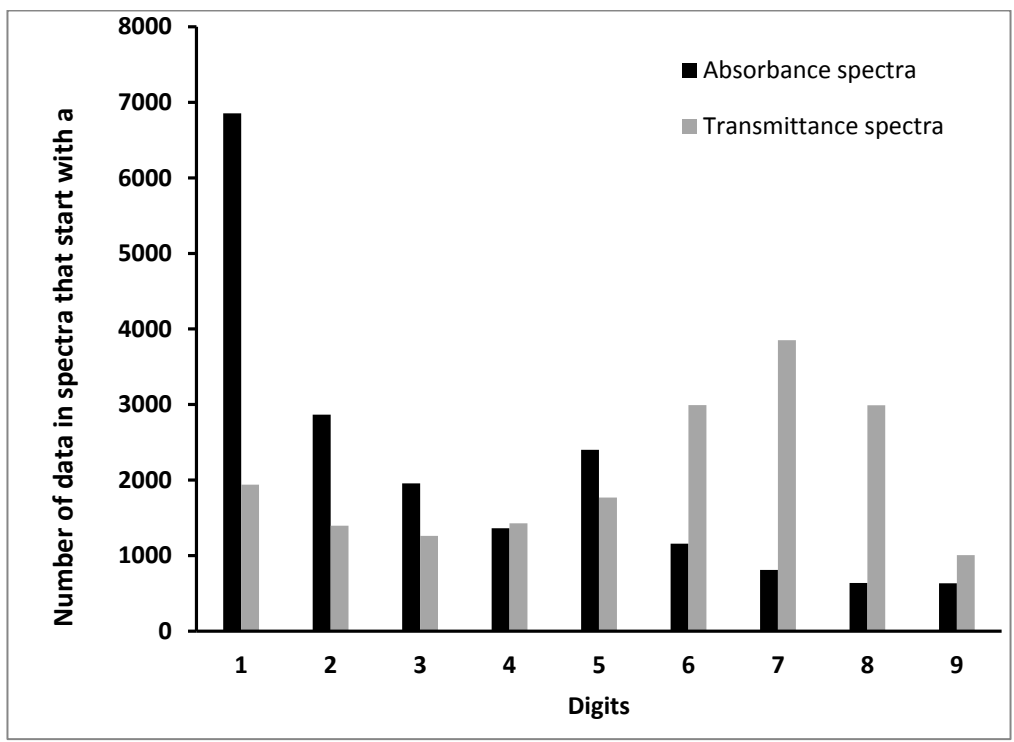

**Fig. 1**. Observed distribution of leading digits in the absorbance and transmittance spectra of polymers.

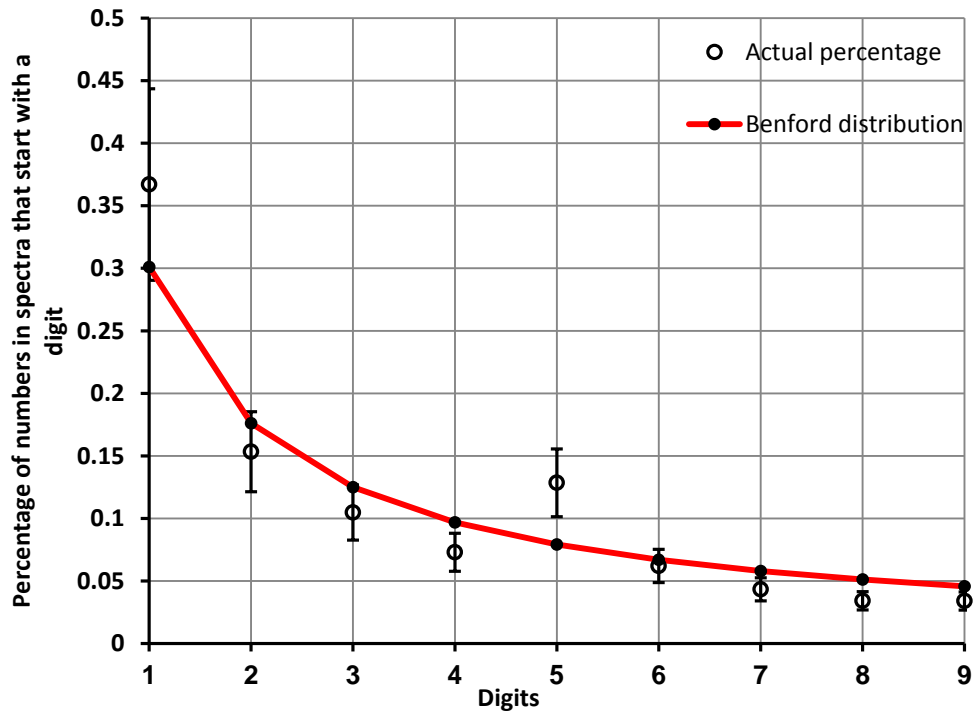

**Fig. 2**. Percentage of leading numbers in the absorbance spectra of polymers. Red solid line represents the Benford distribution. The scale bars demonstrate the standard deviation.

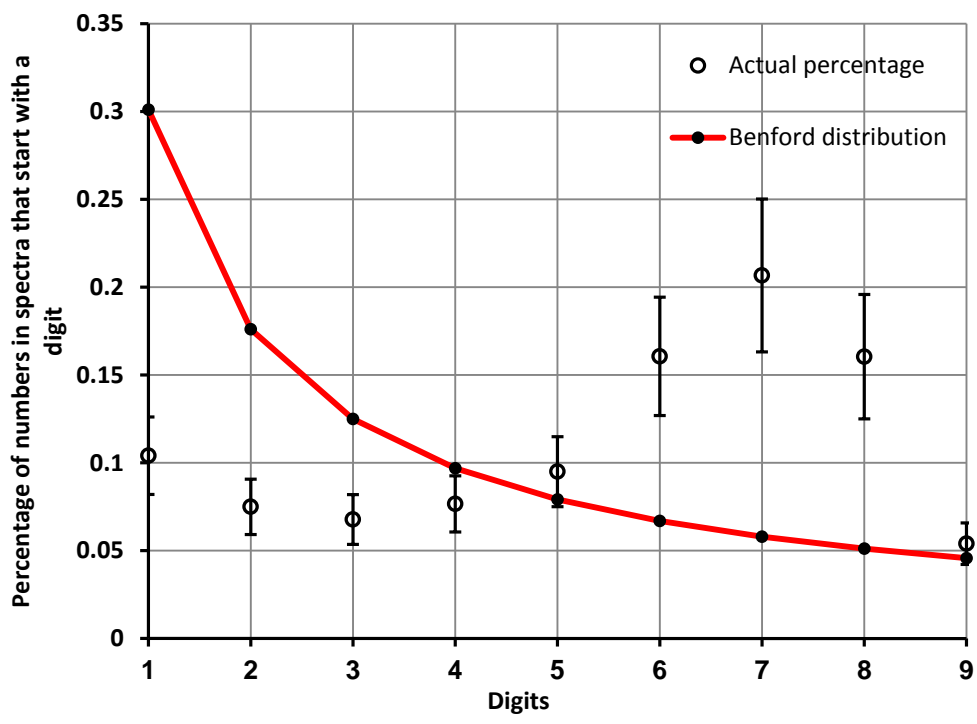

**Fig. 3**. Percentage of leading numbers in the transmittance spectra of polymers. Red solid line represents the Benford distribution. The scale bars demonstrate the standard deviation.

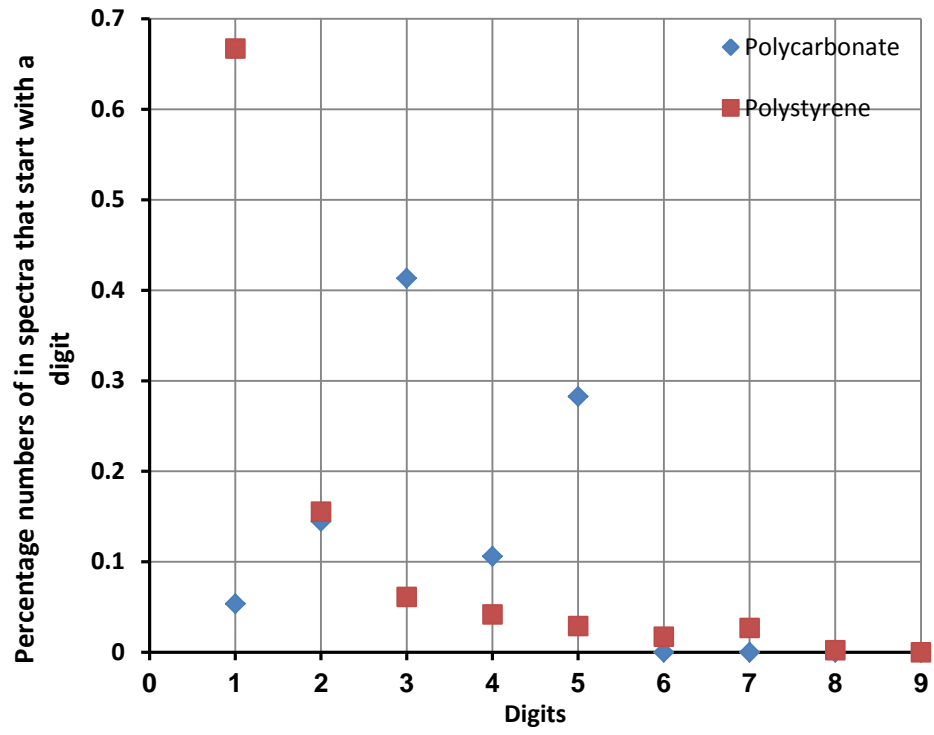

**Fig. 4**. Percentage of leading numbers in the absorbance spectra of polycarbonate (blue diamonds) and polystyrene (red squares). No Benford distribution is recognized.